\documentclass[twocolumn,pra,showpacs]{revtex4}
\usepackage{amsmath,amssymb,latexsym}

\usepackage{mathptmx}
\usepackage{draftcopy}
\newcommand{\hA}{\text{A}}
\newcommand{\hB}{\text{B}}
\newcommand{\hC}{\text{C}}
\newcommand{\hD}{\text{D}}
\newcommand{\hE}{\text{E}}
\newcommand{\hF}{\text{F}}

\begin{document}
\title{No information flow using statistical fluctuations, and quantum
  cryptography}

\author{Jan-\AA ke Larsson} \affiliation{Matematiska Institutionen,
  Link\"opings Universitet, SE-581 83 Link\"oping, Sweden}\medskip

\begin{abstract}
  The communication protocol of Home and Whitaker [Phys.\ Rev.~A {\bf
    67}, 022306 (2003)] is examined in some detail, and found to work
  equally well using a separable state.  The protocol is in fact
  completely classical, based on simple post-selection of suitable
  experimental runs. The quantum cryptography protocol proposed in the
  same publication is also examined, and is found to indeed need
  \emph{quantum} properties for the system to be secure. However, the
  security test proposed in the mentioned paper is found to be
  insufficient, and a modification is proposed here that will ensure
  security.
\end{abstract}
\maketitle

\section{Introduction}

\label{sec:introduction}

Information transfer using quantum entanglement is a subject of great
interest presently. Quantum teleportation
\cite{BBCJPW,BPMEWZ,boschi98:_exper_einst_podol_rosen,NKL} is one of
the more prominent applications, although it has caused some debate
about what is teleported or not \cite{KKJ,quant-ph/0304158}, and on
the relation to nonlocality and inseparability
\cite{popescu94,gisin96:_nonloc,horodecki99:_gener,zukowski00:_bell,clifton01}.
Since quantum entanglement is one of the important resources in
quantum information theory, the interest in these issues is not
surprising. One application is that of quantum computers which would
have a big impact on our world when/if one is actually built. One more
immediate application is quantum cryptography
\cite{BB84,Ekert91,GRTZ}, and one proposal of a quantum-cryptographic
protocol will be discussed below.

First we will look at a quantum communication setup presented in
\cite{HomeWhit}, that uses the experimental setup but not the
communication protocol of the quantum teleportation experiment. The
protocol of \cite{HomeWhit} (to be described briefly below) is
intended to communicate apparatus settings from Alice to Bob without
transmitting the settings on the classical channel that connects Alice
and Bob. This procedure is then extended in \cite{HomeWhit} to make a
quantum cryptography protocol. Here, these two will be critically
examined in order.

The communication protocol is as follows: Alice and Bob share one half
each of many copies of a maximally entangled 2-part spin-$\tfrac12$
state indexed 2 and 3, and Alice has an additional particle at her
disposal indexed 1, so that the total state is
\begin{equation}
    \label{eq:1}
    \left|\Psi_{123}\right\rangle=
    \big(a\left|\uparrow_1\right\rangle
    +b\left|\downarrow_1\right\rangle\big)
    \tfrac1{\sqrt2}
    \big(\left|\uparrow_2\downarrow_3\right\rangle
    -\left|\downarrow_2\uparrow_3\right\rangle\big),
\end{equation}
using the obvious notation for eigenstates of $s_{zn}$ (and we will
denote the measurement results $S_{zn}$ below). The coefficients $a$
and $b$ are chosen to be real here, so that $a^2+b^2=1$. Should one
want to use complex coefficients, each occurrence of $a^2$ and $b^2$
below should be exchanged for $|a|^2$ and $|b|^2$, respectively.  An
alternative way to write the state is
\begin{equation}
  \begin{split}
    \label{eq:2}
    \left|\Psi_{123}\right\rangle=\tfrac12\Big\{&
    \left|\Psi_{12}^+\right\rangle
    \big(-a\left|\uparrow_3\right\rangle
    +b\left|\downarrow_3\right\rangle\big)\\
    &+\left|\Psi_{12}^-\right\rangle
    \big(-a\left|\uparrow_3\right\rangle
    -b\left|\downarrow_3\right\rangle\big)\\
    &+\left|\Phi_{12}^+\right\rangle
    \big(-b\left|\uparrow_3\right\rangle
    +a\left|\downarrow_3\right\rangle\big)\\
    &+\left|\Phi_{12}^-\right\rangle
    \big(+b\left|\uparrow_3\right\rangle
    +a\left|\downarrow_3\right\rangle\big)
    \Big\},
  \end{split}
\end{equation}
where we have used the Bell basis
\begin{equation}
  \begin{split}
    \label{eq:3}
    \left|\Psi_{12}^+\right\rangle
    &=\tfrac1{\sqrt2}\big(\left|\uparrow_1\downarrow_2\right\rangle
    +\left|\downarrow_1\uparrow_2\right\rangle\big)\\
    \left|\Psi_{12}^-\right\rangle
    &=\tfrac1{\sqrt2}\big(\left|\uparrow_1\downarrow_2\right\rangle
    -\left|\downarrow_1\uparrow_2\right\rangle\big)\\
    \left|\Phi_{12}^+\right\rangle
    &=\tfrac1{\sqrt2}\big(\left|\uparrow_1\uparrow_2\right\rangle
    +\left|\downarrow_1\downarrow_2\right\rangle\big)\\
    \left|\Phi_{12}^-\right\rangle
    &=\tfrac1{\sqrt2}\big(\left|\uparrow_1\uparrow_2\right\rangle
    -\left|\downarrow_1\downarrow_2\right\rangle\big).
  \end{split}
\end{equation}
Each run of the protocol uses four of these trios as follows.

Alice performs either (a) four Bell-state measurements on her two
particles 1 and 2 or (b) four measurements of $S_{z1}$ and $S_{z2}$.
In either case, if Alice should happen to get \emph{different}
measurement results on her side in all four measurements [denoted
``criterion Q'' in Ref.~\cite{HomeWhit}], she announces ``OK'' to Bob
on the classical channel. Bob, on the other hand, always performs four
measurements of $S_{z3}$ and calculates the sum $S_{z3\text{t}}$ of
these results. If he received ``OK'' on the classical channel,
the two possibilities of settings at Alice give different probability
distributions of $S_{z3\text{t}}$.

If Alice's setting was (a), Bob has the probability $a^2$ of getting
$+\hbar$ for two of the four particles in the group, and the
probability $b^2$ of getting $+\hbar$ for the other two particles.
This is easily seen in Eq.~(\ref{eq:2}), and implies the following for
the probability distribution of the sum,
\begin{eqnarray}
    &P(S_{z3\text{t}}=2\hbar)=P(S_{z3\text{t}}=-2\hbar)
    =a^4b^4&\notag\\
    \label{eq:4}
    &P(S_{z3\text{t}}=\hbar)=P(S_{z3\text{t}}=-\hbar)
    =2a^2b^2(a^4+b^4)&\\
    &P(S_{z3\text{t}}=0)=a^8+4a^4b^4+b^8.&\notag
\end{eqnarray}
Looking at expectation values, we see that
\begin{equation}
  \label{eq:5}
  \langle S_{z3\text{t}}\rangle=0;
  \quad
  \langle S_{z3\text{t}}^2\rangle=4\hbar^2 a^2b^2,
\end{equation}
where the latter is zero only if $a$ or $b$ is zero. 

If Alice's setting was (b) instead, inspection in Eq.~(\ref{eq:1})
shows that Bob will get two each of ``up'' and ``down'', or in other
words that
\begin{equation}
  \label{eq:6}
  \begin{split}
    P(S_{z3\text{t}}=2\hbar)&=
    P(S_{z3\text{t}}=\hbar)=
    P(S_{z3\text{t}}=-\hbar)\\
    &=P(S_{z3\text{t}}=-2\hbar)=0;\quad
    P(S_{z3\text{t}}=0)=1,
  \end{split}
\end{equation}
so
\begin{equation}
  \label{eq:7}
  \langle S_{z3\text{t}}\rangle=0;
  \quad
  \langle S_{z3\text{t}}^2\rangle=0.
\end{equation}

Bob now checks if the value of $S_{z3\text{t}}$ he received is zero or
nonzero. In the case his measured sum is nonzero he knows with
certainty that Alice's setting was (a), in the ideal case, since it is
obvious in Eq.~(\ref{eq:6}) that nonzero values only occur for setting
(a). If the measured sum is zero, the situation is different. We have
\begin{equation}
  \label{eq:8}
  P\Big(S_{z3\text{t}}=0\,\Big|\,\text{setting (b)}\Big)
  =1,
\end{equation}
while
\begin{equation}
  \label{eq:9}
  P\Big(S_{z3\text{t}}=0\,\Big|\,\text{setting (a)}\Big)
  =a^8+4a^4b^4+b^8.
\end{equation}
To distinguish the setting (a) from the setting (b), one wants the
error probability in Eq.~(\ref{eq:9}) to be as small as possible. It
is minimized if $a=b=1/\sqrt2$, for which it is $3/8= 0.375$.  If the
protocol is modified to use $4N$ triads rather than four, this
probability can be made arbitrarily small. The protocol will indeed
indicate to Bob which setting Alice has used, even though the ``OK''
that Alice sent on the classical channel seems to have nothing to do
with the setting she chose.

\section{A separable-state implementation}

\label{sec:separ-state-impl}

It is noted in Ref.~\cite{HomeWhit} that maximal entanglement is not
necessary for this scheme, but it is argued that entanglement does
play a role. Let us look at this claim closer. Suppose that the state
used is not that of Eq.~(\ref{eq:1})) but the separable
mixed state
\begin{equation}
  \begin{split}
    \label{eq:10}
    \rho_{123}=&
    \big(a^2\left|\uparrow_1\right\rangle\left\langle\uparrow_1\right|
    +b^2\left|\downarrow_1\right\rangle\left\langle\downarrow_1\right|\big)
    \\&\otimes\tfrac12
    \big(\left|\uparrow_2\downarrow_3\right\rangle
    \left\langle\uparrow_2\downarrow_3\right|
    +\left|\downarrow_2\uparrow_3\right\rangle
    \left\langle\downarrow_2\uparrow_3\right|\big).
  \end{split}
\end{equation}
Rewriting this in the Bell basis yields the expression
\begin{equation}
  \begin{split}
    \label{eq:11}
    \rho_{123}=&\tfrac1{4}\Big[
    \big(
    \left|\Psi_{12}^+\right\rangle
    \left\langle\Psi_{12}^+\right|
    +\left|\Psi_{12}^-\right\rangle
    \left\langle\Psi_{12}^-\right|\big)
    \\&\qquad \otimes
    \big(a^2\left|\uparrow_3\right\rangle
    \left\langle\uparrow_3\right|
    +b^2\left|\downarrow_3\right\rangle
    \left\langle\downarrow_3\right|\big)\\
    &\quad+\big(
    \left|\Phi_{12}^+\right\rangle
    \left\langle\Phi_{12}^+\right|
    +\left|\Phi_{12}^-\right\rangle
    \left\langle\Phi_{12}^-\right|\big)
    \\&\qquad \otimes
    \big(b^2\left|\uparrow_3\right\rangle
    \left\langle\uparrow_3\right|
    +a^2\left|\downarrow_3\right\rangle
    \left\langle\downarrow_3\right|\big)    
    \Big]\\
    &+\tfrac1{4}\Big[\big(
    \left|\Psi_{12}^+\right\rangle
    \left\langle\Psi_{12}^-\right|
    +\left|\Psi_{12}^-\right\rangle
    \left\langle\Psi_{12}^+\right|\big)
    \\&\qquad \otimes
    \big(a^2\left|\uparrow_3\right\rangle
    \left\langle\uparrow_3\right|
    -b^2\left|\downarrow_3\right\rangle
    \left\langle\downarrow_3\right|\big)\\
    &\quad-\big(
    \left|\Phi_{12}^+\right\rangle
    \left\langle\Phi_{12}^-\right|
    +\left|\Phi_{12}^-\right\rangle
    \left\langle\Phi_{12}^+\right|\big)
    \\&\qquad \otimes
    \big(b^2\left|\uparrow_3\right\rangle
    \left\langle\uparrow_3\right|
    -a^2\left|\downarrow_3\right\rangle
    \left\langle\downarrow_3\right|\big)
    \Big],
  \end{split}
\end{equation}
where the first parenthesis contains the diagonal elements. The
question now is what happens when using the above protocol on
$\rho_{123}$. Let us assume that Alice has received four different
measurement results [criterion Q] and announced ``OK'' on the
classical channel.

If Alice's setting was (a), Bob has the probability $a^2$ of getting
$+\hbar$ for two of the four particles in the group, and the
probability $b^2$ of getting $+\hbar$ for the other two particles.
This is easily seen in Eq.~(\ref{eq:11}), from this follows that the
probability distribution of the sum is
\begin{eqnarray}
    &P(S_{z3\text{t}}=2\hbar)=P(S_{z3\text{t}}=-2\hbar)=a^4b^4&\notag\\
    \label{eq:12}
    &P(S_{z3\text{t}}=\hbar)=P(S_{z3\text{t}}=-\hbar)
    =2a^2b^2(a^4+b^4)&\\
    &P(S_{z3\text{t}}=0)=a^8+4a^4b^4+b^8.&\notag
\end{eqnarray}
Looking at expectation values, we see that
\begin{equation}
  \label{eq:13}
  \langle S_{z3\text{t}}\rangle=0;
  \quad
  \langle S_{z3\text{t}}^2\rangle=4\hbar^2 a^2b^2,
\end{equation}
where the latter is zero only if $a$ or $b$ is zero. 

If Alice's setting was (b) instead, inspection in Eq.~(\ref{eq:10})
shows that

\begin{equation}
  \label{eq:14}
  \begin{split}
    P(S_{z3\text{t}}=2\hbar)&=
    P(S_{z3\text{t}}=\hbar)=
    P(S_{z3\text{t}}=-\hbar)\\
    &=P(S_{z3\text{t}}=-2\hbar)=0;\quad
    P(S_{z3\text{t}}=0)=1,
  \end{split}
\end{equation}
so
\begin{equation}
  \label{eq:15}
  \langle S_{z3\text{t}}\rangle=0;
  \quad
  \langle S_{z3\text{t}}^2\rangle=0.
\end{equation}
This is exactly the same as Eqs.~(\ref{eq:4})--(\ref{eq:7}), so the
conclusion is the same. There is another way to see this because,
with $a$ and $b$ possibly complex,
\begin{equation}
  \begin{split}
    \label{eq:16}
    \left|\Psi_{123}\right\rangle&\left\langle\Psi_{123}\right|=
    \big(a\left|\uparrow_1\right\rangle
    +b\left|\downarrow_1\right\rangle\big)
    \big(\bar{a}\left\langle\uparrow_1\right|
    +\bar{b}\left\langle\downarrow_1\right|\big)
    \\&\qquad\otimes\tfrac12
    \big(\left|\uparrow_2\downarrow_3\right\rangle
    -\left|\downarrow_2\uparrow_3\right\rangle\big)
    \big(\left\langle\uparrow_2\downarrow_3\right|
    -\left\langle\downarrow_2\uparrow_3\right|\big)
    \\
    &=\rho_{123}-
    \big(a\left|\uparrow_1\right\rangle
    +b\left|\downarrow_1\right\rangle\big)
    \big(\bar{a}\left\langle\uparrow_1\right|
    +\bar{b}\left\langle\downarrow_1\right|\big)
    \\&\qquad\otimes\tfrac12
    \big(\left|\uparrow_2\downarrow_3\right\rangle
    \left\langle\downarrow_2\uparrow_3\right|
    +\left|\downarrow_2\uparrow_3\right\rangle
    \left\langle\uparrow_2\downarrow_3\right|\big)
    \\&\quad+\big(
    a\bar{b}\left|\uparrow_1\right\rangle\left\langle\downarrow_1\right|
    +\bar{a}b\left|\downarrow_1\right\rangle\left\langle\uparrow_1\right|\big)
    \\&\quad\quad\otimes\tfrac12
    \big(\left|\uparrow_2\downarrow_3\right\rangle
    \left\langle\uparrow_2\downarrow_3\right|
    -\left|\downarrow_2\uparrow_3\right\rangle
    \left\langle\downarrow_2\uparrow_3\right|\big).
  \end{split}
\end{equation}
Note that the last two terms are off-diagonal only. Writing
$\left|\Psi_{123}\right\rangle\left\langle\Psi_{123}\right|$ in the
Bell basis on particles 1 and 2 will yield a full matrix in the
density representation, consisting of 64 elements. Let us not write
that large expression here, but simply note that
\begin{equation}
  \begin{split}
    \label{eq:17}
    \left|\Psi_{123}\right\rangle\left\langle\Psi_{123}\right|=&
    \tfrac1{4}\Big[
    \big(
    \left|\Psi_{12}^+\right\rangle
    \left\langle\Psi_{12}^+\right|
    +\left|\Psi_{12}^-\right\rangle
    \left\langle\Psi_{12}^-\right|\big)
    \\&\qquad \otimes
    \big(a^2\left|\uparrow_3\right\rangle
    \left\langle\uparrow_3\right|
    +b^2\left|\downarrow_3\right\rangle
    \left\langle\downarrow_3\right|\big)\\
    &\quad+\big(
    \left|\Phi_{12}^+\right\rangle
    \left\langle\Phi_{12}^+\right|
    +\left|\Phi_{12}^-\right\rangle
    \left\langle\Phi_{12}^-\right|\big)
    \\&\qquad \otimes
    \big(a^2\left|\downarrow_3\right\rangle
    \left\langle\downarrow_3\right|
    +b^2\left|\uparrow_3\right\rangle
    \left\langle\uparrow_3\right|\big)    
    \Big]\\
    &+\Big[\text{off-diagonal terms}\Big]\\
    =&\rho_{123}+\Big[\text{off-diagonal terms}\Big].
  \end{split}
\end{equation}
So, it is clear from Eq.~(\ref{eq:17}) that for (a) Bell-state
measurements at Alice and $S_{z3}$ measurements at Bob, the two states
will yield the same statistics. Similarly, it is clear from
Eq.~(\ref{eq:16}) that for (b) measurements of $S_{zn}$ at Alice and
Bob, the two states will yield the same statistics. The densities have
equal diagonal elements \emph{in both expansions}, and are therefore
impossible to distinguish if Alice and Bob are only allowed to use
measurements (a) or (b), and measurement of $S_{z3}$, respectively. A
discussion of the role of the off-diagonal elements in regular quantum
teleportation can be found in Ref.~\cite{KKJ}.  Of course, the
off-diagonal elements will play a role if Alice and Bob are allowed
other measurements than the ones singled out above.  Especially, only
allowing Bob measurements of $S_{z3}$ is a severe restriction, but we
will investigate this further in
Section~\ref{sec:quantum-cryptography} below.

However, for the proposed protocol, the \emph{separable} mixed state
$\rho_{123}$ will produce the same results as
$\left|\Psi_{123}\right\rangle\left\langle\Psi_{123}\right|$. In fact,
the protocol does not use any specifically \emph{quantum} properties
of the system, it uses classical post\-selection to obtain the desired
statistics. Post\-selection \emph{is} known to sometimes give
unexpected results (see, e.g., \cite{Jalar99c}).

\section{A coin-toss implementation}

\label{sec:coin-toss-impl}

It is not difficult to implement this protocol using a number of
classical unbiased and biased coin tosses. To make it look like the
above setup, we will use three coins $c_1$, $c_2$ and $c_3$ that are
divided so that Alice can read off the results from coins $c_1$ and
$c_2$ while Bob can read the result of coin $c_3$.

Coin $c_1$ is biased so that
\begin{equation}
  \label{eq:18}
  P(c_1=+1)=a^2;\quad P(c_1=-1)=b^2,
\end{equation}
where we have used $+1$ to denote ``heads'' and $-1$ to denote
``tails''.  Coin $c_2$ is fair, i.e.,
\begin{equation}
  \label{eq:19}
  P(c_2=+1)=\tfrac12;\quad P(c_2=-1)=\tfrac12,
\end{equation}
and coin $c_3$ always gives the opposite result to coin $c_2$. This
can be implemented using \emph{one} coin toss at the ``source of the
$c_2$-$c_3$ pair,'' communicating the result to Alice, and the
opposite result to Bob, both on a classical channel. In addition, we
will need a fair coin $c_4$ that Alice will use in case (a) below. An
important comment to make is that the coins $c_1$, $c_2$, and $c_4$
should be independent.

The protocol proceeds as above, with groups of four tosses of the
coins $c_1$--$c_4$. A ``measurement'' consists of reading off the
result of a coin toss. Alice reads either (a) four results of $c_4$
and the product $c_1c_2$ [which, in a way, corresponds to the
Bell-state measurement used previously], or (b) four results of $c_1$
and $c_2$. In either case, if Alice should happen to get four
\emph{different} results on her side [criterion Q], she announces
``OK'' to Bob on the classical channel. Bob, on the other hand, always
reads off $c_3$, and calculates the sum $c_{3\text{t}}$ of the
results. If he received ``OK'' on the classical channel, the two
possibilities of settings at Alice give different conditional
probability distributions of $c_{3\text{t}}$.

The case when Alice used setting (b) is trivial, since Bob will
receive two each of $+1$ and $-1$, so the probability distribution
reads
\begin{equation}
  \label{eq:20}
  \begin{split}
    P(c_{3\text{t}}=2)&=
    P(c_{3\text{t}}=1)=
    P(c_{3\text{t}}=-1)\\
    &=P(c_{3\text{t}}=-2)=0;\quad
    P(c_{3\text{t}}=0)=1,
  \end{split}
\end{equation}
and the classical expectations are
\begin{equation}
  \label{eq:21}
  E( c_{3\text{t}})=0;
  \quad
  E( c_{3\text{t}}^2)=0.
\end{equation}

The case when Alice's setting was (a) is a more complicated and, for
clarity, let us do the calculation explicitly. Since the coins $c_1$
and $c_2$ are independent and $c_2$ is fair, 
\begin{equation}
  \label{eq:22}
  P(c_1c_2=-1)=\tfrac12,
\end{equation}
and since $c_3=-c_2$ we have
\begin{equation}
  \label{eq:23}
  \begin{split}
    P(c_3=-1\cap c_1c_2=&-1)
    =P(c_2=+1\cap c_1=-1)\\
    &=P(c_2=+1)P(c_1=-1)=\tfrac12 b^2.
  \end{split}
\end{equation}
Thus
\begin{subequations}
\begin{equation}
  \label{eq:24}
  P(c_3=-1|c_1c_2=-1)=\frac{P(c_3=-1\cap c_1c_2=-1)}{
    P(c_1c_2=-1)}=b^2,
\end{equation}
and from that follows
\begin{equation}
  \label{eq:25}
  P(c_3=+1|c_1c_2=-1)=a^2.
\end{equation}
\end{subequations}
From a similar calculation we obtain
\begin{equation}
  \label{eq:26}
  \begin{split}
    P(c_3=-1|c_1c_2=+1)=a^2,\\
    P(c_3=+1|c_1c_2=+1)=b^2.\\
  \end{split}
\end{equation}
An ``OK'' from Alice means that the result of $c_1c_2$ was $+1$ twice
and $-1$ twice. This implies that the probability distribution of the
sum $c_{3\text{t}}$ must be
\begin{eqnarray}
    &P(c_{3\text{t}}=2)=P(c_{3\text{t}}=-2)=a^4b^4&\notag\\
    \label{eq:27}
    &P(c_{3\text{t}}=1)=P(c_{3\text{t}}=-1)
    =2a^2b^2(a^4+b^4)&\\
    &P(c_{3\text{t}}=0)=a^8+4a^4b^4+b^8.&\notag
\end{eqnarray}
Looking at expectation values, we see that
\begin{equation}
  \label{eq:28}
  E( c_{3\text{t}})=0;
  \quad
  E( c_{3\text{t}}^2)=4 a^2b^2,
\end{equation}
where the latter is zero only if $a$ or $b$ is zero. 

This is the same statistics as obtained before (modulo the
measurement-result labels $\pm\hbar$), so this completely classical
scheme implements the protocol just as well as the previous two quantum
systems.

\section{Quantum cryptography}

\label{sec:quantum-cryptography}

The second issue in \cite{HomeWhit} is to provide a quantum
cryptography protocol based on the above procedure. As is usual, Alice
and Bob are assumed to have an open but unjammable classical channel
to communicate on, and a quantum channel that in this case consists of
the common source emitting particles 2 and 3. Alice has two random
sequences of bits, one that provides the \emph{raw key} to be
transmitted to Bob over the quantum channel, and another that decides
the \emph{encoding} to be used. Bob has a third random bit sequence
that decides the \emph{decoding} he will use.

Note that the described use of three different random bit sequences is
very similar to the BB84 \cite{BB84} protocol. The difference is that
the present setup uses a source of entangled pairs of qubits, much as
in Ekert quantum cryptography \cite{Ekert91} but, as we will see
shortly, without the Bell inequality test. Another difference is the
usage of several pairs of qubits for transmission of a bit in the key,
as described below.

For each bit with the value 1 in the raw key Alice makes (a) $4N$
Bell-state measurements, and for each bit with the value 0 she uses
the encoding bit-sequence to determine which to perform of (b) $4N$
measurements of $S_{z1}$ and $S_{z2}$, or (c) $4N$ measurements of
$S_{x1}$ and $S_{x2}$. Bob, who knows nothing of Alice's two bit
sequences, uses his decoding bit sequence to determine which to
perform of (b') $4N$ measurements of $S_{z3}$ or (c') $4N$
measurements of $S_{x3}$.

Given that Alice receives the four possible results $N$ times each
[criterion Q], she announces the encoding to Bob (but not the key
bit), i.e., which setting of (b) or (c) she used. If she happened to
use setting (a), the encoding bit will make her announce one of (b) or
(c) to Bob, randomly with equal probability. This means that Alice
transmits the encoding bit to Bob, but not the raw key bit; she does
not transmit any information about the raw key over the classical
channel.

Bob discards data where his setting was (b') and Alice announced (c),
and where his setting was (c') and Alice announced (b). For the
remaining runs, when he used setting (b') he can determine whether
Alice's setting was (a) or (b) by the earlier protocol, and similarly
when he used setting (c') he can determine whether Alice's setting was
(a) or (c). He can now determine the bits of the raw key for the
experimental runs that remain after the above filtering. He also
communicates which runs he is using to Alice, but neither the settings
nor the resulting bit. The remaining bit sequence (the sifted key)
will now be equal at Alice and Bob, or at least as equal as possible,
see below. They have established a key to use in their cryptographic
scheme.

There will be some noise in the sifted key even in the ideal case,
because Bob cannot with certainty say, for example, whether Alice used
(a) or (b). If Bob receives a nonzero result in his measurement of
$S_{z3\text{t}}$, he knows that Alice's setting was (a) but if Bob
receives a zero result, the probability that Alice used setting (b) is
larger than $\tfrac12$ but there is a nonzero probability that the
setting was (a). This probability will depend on $a$, $b$ and $N$ and
tend to zero as $N$ tends to infinity. In a real implementation one
has to choose $N$ finite, otherwise the key rate will be zero. For
example, when $a=b$ and $N=1$,

\begin{center}
  \begin{tabular}{c c | c}
    \multicolumn{3}{c}{Table 1}\\
    \multicolumn{3}{c}{Normal operation: $P($OK$)=\tfrac6{64}$
      \vspace{3pt}}\\
    Alice's bit & Bob's bit & Probability\\
    \hline
    0 & 0 & 1\\
    0 & 1 & 0\\
    1 & 0 & $\tfrac38$\\
    1 & 1 & $\tfrac58$\\
  \end{tabular}
\end{center}

The visible effect will be that some ones in Alice's bit-sequence will
arrive as the value zero at Bob. In the ideal case, no zeros will
become one, so evidently, Bob's copy of the sifted key will have
slightly more zeros than Alice's copy.

Comparing this cryptographic scheme with the previously described
communication scheme, the set of possible measurement setups is
extended so that the off-diagonal terms in the expansion of
$\left|\Psi_{123}\right\rangle\left\langle\Psi_{123}\right|$ come into
play, making that particular state or another entangled state a
requirement for the quantum cryptography scheme.  And the whole idea
of quantum cryptography is to use specifically \emph{quantum}
properties of a system in such a way that eavesdropping always will be
detectable. It is noted in Ref.~\cite{HomeWhit} that if
$\left|\Psi_{123}\right\rangle\left\langle\Psi_{123}\right|$ is used,
and the eavesdropper Eve makes $4N$ measurements of either $S_{z3}$ or
$S_{x3}$ at random, she will be detected. In the case $a=b$ and $N=1$,
we have
\begin{center}
  \begin{tabular}{c c c | c}
    \multicolumn{4}{c}{Table 2}\\
    \multicolumn{4}{c}{Eve is listening: $P($OK$)=\tfrac6{64}$
      \vspace{3pt}}\\
    Alice's bit & Eve's basis & Bob's bit & Probability\\
    \hline
    0 & correct & 0 & 1\\
    0 & correct & 1 & 0\\
    0 & incorrect & 0 & $\tfrac38$\\
    0 & incorrect & 1 & $\tfrac58$\\
    0 & (mean) & 0 & $\tfrac{11}{16}$\\
    0 & (mean) & 1 & $\tfrac5{16}$\\
    1 & either & 0 & $\tfrac38$\\
    1 & either & 1 & $\tfrac58$\\
  \end{tabular}
  \vspace*{0pt}
\end{center}
Apparently, there will be extra noise, but only in the zeros. In
comparison to the BB84 protocol, the situation is as follows:

\begin{center}
  \begin{tabular}{c c | c c}
    \multicolumn{4}{c}{Table 3}\\
    &&BB84&HW\cite{HomeWhit}\\
    Alice's bit & Eve is & $P($error$)$ & $P($error$)$\\
    \hline
    0 & absent & 0 & 0\\
    1 & absent & 0 & $\tfrac38$\\
    0 & present & $\tfrac14$ & $\tfrac5{16}$\\
    1 & present & $\tfrac14$ & $\tfrac38$
  \end{tabular}
  \vspace*{0pt}
\end{center}

The performance of BB84 is better than the protocol of Home and
Whitaker \cite{HomeWhit} when Eve is absent, while the figures are
comparable when Eve is present.  Of course, the protocol of
\cite{HomeWhit} uses more qubits per sifted key bit than BB84. Both
protocols will reject half the data outright (when the ``settings''
disagree at Alice and Bob), but in addition, the protocol of
\cite{HomeWhit} uses four qubits for each raw key bit and of these
only 6/64 will yield a bit in the sifted key, that is, when Alice
announces ``OK'' on the classical channel. Also there is another
problem, which we will turn to now.

\section{A coherent attack}

\label{sec:coherent-attack}

If the source is in an insecure location, or if Eve has access to both
the quantum channel going from the source to Alice and that going to
Bob, she can replace the source with her own.  Eve can of course
replace the source emitting the entangled state with a source randomly
emitting either $4N$ copies of the mixed state
\begin{subequations}
  \begin{equation}
    \tfrac12
    \big(\left|\uparrow_{z2}\downarrow_{z3}\right\rangle
    \left\langle\uparrow_{z2}\downarrow_{z3}\right|
    +\left|\downarrow_{z2}\uparrow_{z3}\right\rangle
    \left\langle\downarrow_{z2}\uparrow_{z3}\right|\big),
  \end{equation}
  or $4N$ copies of
  \begin{equation}
    \label{eq:29}
    \tfrac12
    \big(\left|\uparrow_{x2}\downarrow_{x3}\right\rangle
    \left\langle\uparrow_{x2}\downarrow_{x3}\right|
    +\left|\downarrow_{x2}\uparrow_{x3}\right\rangle
    \left\langle\downarrow_{x2}\uparrow_{x3}\right|\big),
  \end{equation}
\end{subequations}
but that would yield the same statistics as that obtained when Eve is
simply eavesdropping on (one of) the quantum channels. And then, Alice
and Bob can use a statistical test to detect Eve's precense. 

Eve can do something more clever than transmitting $4N$ copies of a
mixed state, but let us now restrict ourselves to the case $N=1$ and
$a=b$ for simplicity. Note that Eve has complete freedom of choosing
what state to send to Alice and Bob, including what \emph{sequence} of
states to send.  She can for instance choose to send the sequence
\begin{equation}
  \begin{split}
    \left|\uparrow_{z2}\downarrow_{z3}\right\rangle
    \left\langle\uparrow_{z2}\downarrow_{z3}\right|,\;
    \left|\uparrow_{z2}\downarrow_{z3}\right\rangle
    \left\langle\uparrow_{z2}\downarrow_{z3}\right|,\\
    \left|\downarrow_{z2}\uparrow_{z3}\right\rangle
    \left\langle\downarrow_{z2}\uparrow_{z3}\right|,\;
    \left|\downarrow_{z2}\uparrow_{z3}\right\rangle
    \left\langle\downarrow_{z2}\uparrow_{z3}\right|,
\end{split}
\end{equation}
which would yield a key bit of zero at Bob if Bob uses the (b')
setting: measurement of $S_{z3\text{tot}}$; in the above sequence, the
sum of the received spins is zero. At Alice, the sequence will yield
``OK'' if Alice uses the (b) setting. In a different notation, Eve
will have sent the state
\begin{equation}
  \left|\hC_{z2}\right\rangle
  =\left|\uparrow_{z2}^1\uparrow_{z2}^2
    \downarrow_{z2}^3\downarrow_{z2}^4\right\rangle
\end{equation}
to Alice (the upper index on the right-hand side denotes the timeslot,
and in hexadecimal $\hC_{16}=1100_2$ perhaps with the name
``qunybble'' \endnote{In computer science a ``nybble'' is half a byte,
  or four bits. Four qubits would then constitute a qunybble.}), and
\begin{equation}
  \left|3_{z3}\right\rangle
  =\left|\downarrow_{z3}^1\downarrow_{z3}^2
    \uparrow_{z3}^3\uparrow_{z3}^4\right\rangle
\end{equation}
to Bob. Obviously, this state is only good if Alice and Bob do not use
(c) and (c'). If they do, the probability of ``OK'' at Alice is 6/64
since the results will be random at Alice, and there will be some
noise in the ``transmitted'' key bit at Bob, since the measurement
results will be random there as well. But Eve has one more ace up her
sleeve: entanglement.
 
Note that
\begin{subequations}
  \begin{equation}
    \label{eq:30}
    \begin{split}
      \left|3_{z3}\right\rangle
      +\left|\hC_{z3}\right\rangle=&\frac12 \Big[
      \Big(\left|0_{x3}\right\rangle+\left|\hF_{x3}\right\rangle\Big)
      +\Big(\left|3_{x3}\right\rangle+\left|\hC_{x3}\right\rangle\Big)\\
      &\quad-\Big(\left|5_{x3}\right\rangle+\left|\hA_{x3}\right\rangle\Big)
      -\Big(\left|9_{x3}\right\rangle+\left|6_{x3}\right\rangle\Big)
      \Big],
  \end{split}
\end{equation}
  \begin{equation}
    \label{eq:31}
    \begin{split}
      \left|5_{z3}\right\rangle
      +\left|\hA_{z3}\right\rangle=&\frac12 \Big[
      \Big(\left|0_{x3}\right\rangle+\left|\hF_{x3}\right\rangle\Big)
      -\Big(\left|3_{x3}\right\rangle+\left|\hC_{x3}\right\rangle\Big)\\
      &\quad+\Big(\left|5_{x3}\right\rangle+\left|\hA_{x3}\right\rangle\Big)
      -\Big(\left|9_{x3}\right\rangle+\left|6_{x3}\right\rangle\Big)
      \Big],
  \end{split}
\end{equation}
and
  \begin{equation}
    \label{eq:32}
    \begin{split}
      \left|9_{z3}\right\rangle
      +\left|6_{z3}\right\rangle=&\frac12 \Big[
      \Big(\left|0_{x3}\right\rangle+\left|\hF_{x3}\right\rangle\Big)
      -\Big(\left|3_{x3}\right\rangle+\left|\hC_{x3}\right\rangle\Big)\\
      &\quad-\Big(\left|5_{x3}\right\rangle+\left|\hA_{x3}\right\rangle\Big)
      +\Big(\left|9_{x3}\right\rangle+\left|6_{x3}\right\rangle\Big)
      \Big].
  \end{split}
\end{equation}
\end{subequations}
Letting $q=\exp(2i\pi/3)$, Eve can choose to transmit the state 
$\big|\psi_{3,00}\big\rangle$, given by
\begin{equation}
  \label{eq:33}
  \begin{split}
    &\sqrt6\big|\psi_{3,00}\big\rangle\\
    =&\Big(\left|3_{z3}\right\rangle+\left|\hC_{z3}\right\rangle\Big)
    +q\Big(\left|5_{z3}\right\rangle +\left|\hA_{z3}\right\rangle\Big)
    +q^2\Big(\left|9_{z3}\right\rangle +\left|6_{z3}\right\rangle\Big)\\
    =&\Big(\left|3_{x3}\right\rangle+\left|\hC_{x3}\right\rangle\Big)
    +q\Big(\left|5_{x3}\right\rangle +\left|\hA_{x3}\right\rangle\Big)
    +q^2\Big(\left|9_{x3}\right\rangle +\left|6_{x3}\right\rangle\Big),
  \end{split}
\end{equation}
for which the result of measuring \emph{either of} the sum
$S_{z3\text{tot}}$ and the sum $S_{x3\text{tot}}$ is zero. The above
is an entangled state, but the entanglement is in the sequence of
particles emitted to Bob instead of the pairwise entanglement between
particles 2 and 3 in the original source.  Bob will always receive the
measurement result zero, irrespective if his setting is (b') or (c');
this particular source is ``tuned'' for zeros. In addition, if Eve
transmits the same state to Alice, or rather the corresponding
$\big|\psi_{2,00}\big\rangle$, Alice will have a higher chance of
getting ``OK'' if she uses the setting (b) or (c), since the results
at particle 2 will be two of ``up'' and two ``down''. The probability
of getting ``OK'' is then $\tfrac14$ (or 16/64) rather than the usual
$6/64$. We have
\begin{center}
  \begin{tabular}{c c c | c}
    \multicolumn{4}{c}{Table 4}\\
    \multicolumn{4}{c}{Eve tunes the source for zeros}\\
    Alice's bit & $P($OK$)$ & Bob's bit & Probability\\
    \hline
    0 & $\tfrac{16}{64}$ & 0 & 1\\
    0 & $\tfrac{16}{64}$ & 1 & 0\\
    1 & $\tfrac6{64}$ & 0 & 1\\
    1 & $\tfrac6{64}$ & 1 & 0\\
  \end{tabular}
  \vspace*{0pt}
\end{center}

Eve cannot use this source only, since only zeros would be transmitted
(correctly) to Bob. But Eve can also tune the source to produce ones;
note that
  \begin{equation}
    \label{eq:34}
    \begin{split}
      \left|0_{z3}\right\rangle -\left|\hF_{z3}\right\rangle=&\frac12 \Big[
      \Big(\left|1_{x3}\right\rangle+\left|\hE_{x3}\right\rangle\Big)
      +\Big(\left|2_{x3}\right\rangle+\left|\hD_{x3}\right\rangle\Big)\\
      &\quad+\Big(\left|4_{x3}\right\rangle+\left|\hB_{x3}\right\rangle\Big)
      +\Big(\left|8_{x3}\right\rangle+\left|7_{x3}\right\rangle\Big)
      \Big],
    \end{split}
  \end{equation}  
while
\begin{subequations}
  \begin{equation}
    \label{eq:35}
    \begin{split}
      \left|1_{z3}\right\rangle
      -\left|\hE_{z3}\right\rangle=&\frac12 \Big[
      -\Big(\left|1_{x3}\right\rangle-\left|\hE_{x3}\right\rangle\Big)
      +\Big(\left|2_{x3}\right\rangle-\left|\hD_{x3}\right\rangle\Big)\\
      &\quad+\Big(\left|4_{x3}\right\rangle-\left|\hB_{x3}\right\rangle\Big)
      +\Big(\left|8_{x3}\right\rangle-\left|7_{x3}\right\rangle\Big)
      \Big],
    \end{split}
  \end{equation}  
  \begin{equation}
    \label{eq:36}
    \begin{split}
      \left|2_{z3}\right\rangle
      -\left|\hD_{z3}\right\rangle=&\frac12 \Big[
      \Big(\left|1_{x3}\right\rangle-\left|\hE_{x3}\right\rangle\Big)
      -\Big(\left|2_{x3}\right\rangle-\left|\hD_{x3}\right\rangle\Big)\\
      &\quad+\Big(\left|4_{x3}\right\rangle-\left|\hB_{x3}\right\rangle\Big)
      +\Big(\left|8_{x3}\right\rangle-\left|7_{x3}\right\rangle\Big)
      \Big],
    \end{split}
  \end{equation}
  \begin{equation}
    \label{eq:37}
    \begin{split}
      \left|4_{z3}\right\rangle
      -\left|\hB_{z3}\right\rangle=&\frac12 \Big[
      \Big(\left|1_{x3}\right\rangle-\left|\hE_{x3}\right\rangle\Big)
      +\Big(\left|2_{x3}\right\rangle-\left|\hD_{x3}\right\rangle\Big)\\
      &\quad-\Big(\left|4_{x3}\right\rangle-\left|\hB_{x3}\right\rangle\Big)
      +\Big(\left|8_{x3}\right\rangle-\left|7_{x3}\right\rangle\Big)
      \Big],
    \end{split}
  \end{equation}
and
  \begin{equation}
    \label{eq:38}
    \begin{split}
      \left|8_{z3}\right\rangle
      -\left|7_{z3}\right\rangle=&\frac12 \Big[
      \Big(\left|1_{x3}\right\rangle-\left|\hE_{x3}\right\rangle\Big)
      +\Big(\left|2_{x3}\right\rangle-\left|\hD_{x3}\right\rangle\Big)\\
      &\quad+\Big(\left|4_{x3}\right\rangle-\left|\hB_{x3}\right\rangle\Big)
      -\Big(\left|8_{x3}\right\rangle-\left|7_{x3}\right\rangle\Big)
      \Big].
    \end{split}
  \end{equation}
 \end{subequations}
So Eve could use the state
$\left|\psi_{3,11}\right\rangle$, where
\begin{equation}
  \label{eq:39}
  \begin{split}
    \sqrt{10}\big|\psi_{3,11}\big\rangle=&
      \Big(\left|0_{z3}\right\rangle-\left|\hF_{z3}\right\rangle\Big)\\
      &+q\Big(\left|1_{z3}\right\rangle+\left|2_{z3}\right\rangle
      +\left|4_{z3}\right\rangle+\left|8_{z3}\right\rangle\Big)\\
      &+q^2\Big(\left|\hE_{z3}\right\rangle+\left|\hD_{z3}\right\rangle
      +\left|\hB_{z3}\right\rangle+\left|6_{z3}\right\rangle\Big)\\
      =&
      \Big(\left|0_{z2}\right\rangle-\left|\hF_{z2}\right\rangle\Big)\\
      &+q\Big(\left|1_{z2}\right\rangle+\left|2_{z2}\right\rangle
      +\left|4_{z2}\right\rangle+\left|8_{z2}\right\rangle\Big)\\
      &+q^2\Big(\left|\hE_{z2}\right\rangle+\left|\hD_{z2}\right\rangle
      +\left|\hB_{z2}\right\rangle+\left|6_{z2}\right\rangle\Big),
  \end{split}
\end{equation}
for which the relevant probabilities are
\begin{center}
  \begin{tabular}{c c c | c}
    \multicolumn{4}{c}{Table 5}\\
    \multicolumn{4}{c}{Eve tunes the source for ones}\\
    Alice's bit & $P($OK$)$ & Bob's bit & Probability\\
    \hline
    0 & 0 & 0 & 0\\
    0 & 0 & 1 & 1\\
    1 & $\tfrac6{64}$ & 0 & 0\\
    1 & $\tfrac6{64}$ & 1 & 1\\
  \end{tabular}
  \vspace*{0pt}
\end{center}

Eve does not want to change the ratio of ones to zeros in the
transmitted key from that of the raw key, since that would enable a
statistical test for her presence.  The rate of OK's should be $6/64$
irrespective of whether Alice has a 0 or a 1 in her raw key. This will
be achieved by letting Eve's source be tuned for zeros with a
probability of $6/16$, and be tuned for ones with a probability of
$10/16$. In quantum language, Eve would tune the source to send
\begin{equation}
  \label{eq:40}
  \tfrac38\big|\psi_{2,00}\psi_{3,00}\big\rangle
  \big\langle\psi_{2,00}\psi_{3,00}\big|
  +\tfrac58\big|\psi_{2,11}\psi_{3,11}\big
  \rangle\big\langle\psi_{2,11}\psi_{3,11}\big|.
\end{equation}
Remarkably, this is the same ratio as is required to make the 16
different bit combinations of the four particles equally probable,
something also desired by Eve, because otherwise, e.g., Bob could test
the statistical properties of his unfiltered data to detect Eve. With
this ratio, the probability of ``OK'' is
\begin{subequations}
\begin{equation}
  \label{eq:41}
  P\Big(\text{OK}\,\Big|\,\text{Raw key bit is 0}\Big)
  =\tfrac6{16}\cdot\tfrac{16}{64}+\tfrac{10}{16}\cdot0=\tfrac6{64}
\end{equation}
and
\begin{equation}
  \label{eq:42}
  P\Big(\text{OK}\,\Big|\,\text{Raw key bit is 1}\Big)
  =\tfrac6{16}\cdot\tfrac{6}{64}+\tfrac{10}{16}\cdot\tfrac{6}{64}=\tfrac6{64}
\end{equation}
\end{subequations}
Zeros are transferred only when the source is tuned for zeros, which
also means that there will be no errors in the zeros, in the ideal
case. However, ones will be ``transferred'' both when the source is
tuned for ones (no errors in the ideal case) and when the source is
tuned for zeros (all errors in the ideal case). With the above
weighting, the rate of errors in the ones will be
\begin{equation}
  \label{eq:43}
  P\Big(\text{Bob gets a 0}\,\Big|\,
  \text{OK}\,,\,\text{Raw key bit is 1}\Big)
  =\tfrac6{16}\cdot1+\tfrac{10}{16}\cdot0=\tfrac38
\end{equation}
We arrive at
\begin{center}
  \begin{tabular}{c c | c}
    \multicolumn{3}{c}{Table 6}\\
    \multicolumn{3}{c}{Eve tunes the source: $P($OK$)=\tfrac6{64}$
      \vspace{3pt}}\\
    Alice's bit & Bob's bit & Probability\\
    \hline
    0 & 0 & 1\\
    0 & 1 & 0\\
    1 & 0 & $\tfrac38$\\
    1 & 1 & $\tfrac58$\\
  \end{tabular}
\end{center}
This table is identical to Table 1, ``Normal operation.'' Eve is
controlling the source, and she knows what values Bob will recieve
when Alice announces ``OK'' on the classical channel. She has, quite
surprisingly, used entanglement to her benefit for a coherent attack
on the four qubits making up a single key bit value. The errors occur
exactly like in the case Eve is absent.

\section{The perfect illusion?}

\label{sec:perfect-illusion}

So Eve has a way to eavesdrop unnoticed on Alice and Bob, at least if
the security test used is the one proposed in
Section~\ref{sec:quantum-cryptography}, estimating the error rate in
the sifted key. But note that with the above tuned source, if Alice
receives a spin sum is equal to zero, Bob will also receive a spin sum
equal to zero, \emph{regardless of the setting they use}. And
(provided they have read this paper) Alice and Bob will by now know to
test this in their system. Is this then enough?

Let us see if Eve can construct a system that obeys the following:
\begin{enumerate}
\renewcommand{\theenumi}{\roman{enumi}}
\renewcommand{\labelenumi}{(\theenumi)}
\item Eve can eavesdrop, or alternatively, control the source so that
  she knows in advance what the results should be for both measurement
  settings
\item if Alice gets spin-sum zero at one setting, then Bob also should
  get spin-sum zero at the same setting
\item the spin-sum at one setting at Bob is statistically independent
  of the spin-sum at the other setting at Alice, and vice versa
\item the 16 different local results are equally probable at either
  setting
\end{enumerate}

It takes some calculation to determine a state with these desired
properties (desired by Eve, that is), but one such state is
\begin{equation}
  \begin{split}
    \label{eq:44}
    &\tfrac9{64}\big|\psi_{2,00}\psi_{3,00}\big\rangle
    \big\langle\psi_{2,00}\psi_{3,00}\big|\\
    &+\tfrac5{64}\big|\alpha_{2,01}\alpha_{3,01}\big
    \rangle\big\langle\alpha_{2,01}\alpha_{3,01}\big|
    +\tfrac5{64}\big|\alpha'_{2,10}\alpha'_{3,10}\big
    \rangle\big\langle\alpha'_{2,10}\alpha'_{3,10}\big|\\
    &+\tfrac5{64}\big|\beta_{2,01}\beta_{3,01}\big\rangle
    \big\langle\beta_{2,01}\beta_{3,01}\big|
    +\tfrac5{64}\big|\beta'_{2,10}\beta'_{3,10}\big\rangle
    \big\langle\beta'_{2,10}\beta'_{3,10}\big|\\
    &+\tfrac5{64}\big|\gamma_{2,01}\gamma_{3,01}\big
    \rangle\big\langle\gamma_{2,01}\gamma_{3,01}\big|
    +\tfrac5{64}\big|\gamma'_{2,10}\gamma'_{3,10}\big
    \rangle\big\langle\gamma'_{2,10}\gamma'_{3,10}\big|\\
    &+\tfrac8{64}\big|\chi_{2,11}\chi_{3,11}\big
    \rangle\big\langle\chi_{2,11}\chi_{3,11}\big|
    +\tfrac8{64}\big|\chi'_{2,11}\chi'_{3,11}\big
    \rangle\big\langle\chi'_{2,11}\chi'_{3,11}\big|\\
    &+\tfrac9{64}\big|\phi_{2,11}\phi_{3,11}\big
    \rangle\big\langle\phi_{2,11}\phi_{3,11}\big|.
  \end{split}
\end{equation}
Here, the first index of the pure states in the pure-state expansion
is the particle index as used before, and the later two indices
indicate which bit value the particular state is tuned for, in the
bases (b) and (c), in order. The included states are
$\big|\psi_{3,00}\big\rangle$ as defined in Eq.~(\ref{eq:33}), and
\begin{subequations}
\begin{equation}
  \label{eq:45}
  \begin{split}
    \left|\alpha_{3,01}\right\rangle=&\frac1{\sqrt2}\Big(
    \left|3_{z3}\right\rangle -\left|\hC_{z3}\right\rangle\Big)\\
    =&\frac1{2\sqrt2}\Big[
    -\Big(\left|1_{x3}\right\rangle-\left|\hE_{x3}\right\rangle\Big)
    -\Big(\left|2_{x3}\right\rangle-\left|\hD_{x3}\right\rangle\Big)\\
    &\quad\quad
    +\Big(\left|4_{x3}\right\rangle-\left|\hB_{x3}\right\rangle\Big)
    +\Big(\left|8_{x3}\right\rangle-\left|7_{x3}\right\rangle\Big)
    \Big],
  \end{split}
\end{equation}
\begin{equation}
  \label{eq:46}
  \begin{split}
    \left|\beta_{3,01}\right\rangle=&\frac1{\sqrt2}\Big(
    \left|5_{z3}\right\rangle -\left|\hA_{z3}\right\rangle\Big)\\
    =&\frac1{2\sqrt2}\Big[
    -\Big(\left|1_{x3}\right\rangle-\left|\hE_{x3}\right\rangle\Big)
    +\Big(\left|2_{x3}\right\rangle-\left|\hD_{x3}\right\rangle\Big)\\
    &\quad\quad
    -\Big(\left|4_{x3}\right\rangle-\left|\hB_{x3}\right\rangle\Big)
    +\Big(\left|8_{x3}\right\rangle-\left|7_{x3}\right\rangle\Big)
    \Big],
  \end{split}
\end{equation}
\begin{equation}
  \label{eq:47}
  \begin{split}
    \left|\gamma_{3,01}\right\rangle=&\frac1{\sqrt2}\Big(
    \left|9_{z3}\right\rangle -\left|6_{z3}\right\rangle\Big)\\
    =&\frac1{2\sqrt2}\Big[
    -\Big(\left|1_{x3}\right\rangle-\left|\hE_{x3}\right\rangle\Big)
    +\Big(\left|2_{x3}\right\rangle-\left|\hD_{x3}\right\rangle\Big)\\
    &\quad\quad
    +\Big(\left|4_{x3}\right\rangle-\left|\hB_{x3}\right\rangle\Big)
    -\Big(\left|8_{x3}\right\rangle-\left|7_{x3}\right\rangle\Big)
    \Big],
  \end{split}
\end{equation}
\end{subequations}
together with their ``mirrored'' counterparts
\begin{subequations}
\begin{equation}
  \label{eq:48}
  \begin{split}
    \left|\alpha'_{3,10}\right\rangle
    =&\frac1{2\sqrt2}\Big[
    -\Big(\left|1_{z3}\right\rangle-\left|\hE_{z3}\right\rangle\Big)
    -\Big(\left|2_{z3}\right\rangle-\left|\hD_{z3}\right\rangle\Big)\\
    &\quad\quad
    +\Big(\left|4_{z3}\right\rangle-\left|\hB_{z3}\right\rangle\Big)
    +\Big(\left|8_{z3}\right\rangle-\left|7_{z3}\right\rangle\Big)
    \Big]\\
    =&\frac1{\sqrt2}\Big(
    \left|3_{x3}\right\rangle -\left|\hC_{x3}\right\rangle\Big),
  \end{split}
\end{equation}
\begin{equation*}
  \vdots\quad.
\end{equation*}
\end{subequations}
We also need the state
\begin{equation}
  \label{eq:49}
  \begin{split}
    \left|\chi_{3,01}\right\rangle=&\frac1{\sqrt2}\Big(
    \left|0_{z3}\right\rangle -\left|\hF_{z3}\right\rangle\Big)\\
    =&\frac1{2\sqrt2}\Big[
    \Big(\left|1_{x3}\right\rangle+\left|\hE_{x3}\right\rangle\Big)
    +\Big(\left|2_{x3}\right\rangle+\left|\hD_{x3}\right\rangle\Big)\\
    &\quad\quad
    +\Big(\left|4_{x3}\right\rangle+\left|\hB_{x3}\right\rangle\Big)
    +\Big(\left|8_{x3}\right\rangle+\left|7_{x3}\right\rangle\Big)
    \Big],
  \end{split}
\end{equation}
mirrored in the same way, and finally,

\begin{equation}
  \label{eq:50}
  \begin{split}
    \left|\phi_{3,11}\right\rangle=&\frac1{2\sqrt2}\Big[
    \Big(\left|1_{z3}\right\rangle-\left|\hE_{z3}\right\rangle\Big)
    +\Big(\left|2_{z3}\right\rangle-\left|\hD_{z3}\right\rangle\Big)\\
    &\quad\quad
    +\Big(\left|4_{z3}\right\rangle-\left|\hB_{z3}\right\rangle\Big)
    +\Big(\left|8_{z3}\right\rangle-\left|7_{z3}\right\rangle\Big)
    \Big]\\
    =&\frac1{2\sqrt2}\Big[
    \Big(\left|1_{x3}\right\rangle-\left|\hE_{x3}\right\rangle\Big)
    +\Big(\left|2_{x3}\right\rangle-\left|\hD_{x3}\right\rangle\Big)\\
    &\quad\quad
    +\Big(\left|4_{x3}\right\rangle-\left|\hB_{x3}\right\rangle\Big)
    +\Big(\left|8_{x3}\right\rangle-\left|7_{x3}\right\rangle\Big)
    \Big].
  \end{split}
\end{equation}
Each of these states show the same properties as the states that were
used in Section~\ref{sec:coherent-attack}, given that Alice and Bob
use the same settings, but show different behaviour when they use
different settings.  For example, if Eve sends
$\left|\beta_{2,01}\beta_{3,01}\right\rangle$ and Alice and Bob both
use the (b) setting (or Alice (a) and Bob (b)), the results will
follow Table 4 because this state is tuned for zeros in this basis. If
Alice and Bob both use the (c) setting (or Alice (a) and Bob (c)), the
results will follow Table 5 because this state is tuned for ones in
this basis. If they use different settings, the spin sum will differ
at the two sites.

The statistics obtained from the state in Eq.~(\ref{eq:44}) follows
the behaviour of the original state exactly, as far as the spin sums
are concerned, and Table 6 gives the key-transmission errors. The
existence of this state shows that only checking the key bits, or
indeed, checking statistical properties of the spin sums at the two
sites at \emph{any} combination of the two settings, will not provide
any security. 

Is the illusion perfect, then? Not at all. For example, if the spin
sum is one of the extreme values at one site, e.g., a result of
$0_{z2}$ or $\hF_{z2}$, there will be no occurrences of spin sum zero
at the other site at the other setting, i.e., none of $3_{x3}$,
$5_{x3}$, $6_{x3}$, $9_{x3}$, $A_{x3}$, and $C_{x3}$ will occur.
Simply put, this is because there is no possibility to make
\begin{equation}
\label{eq:51}
  \begin{split}
    \left|0_{z3}\right\rangle &+\left|\hF_{z3}\right\rangle
    =c_1\Big(\left|3_{x3}\right\rangle+\left|\hC_{x3}\right\rangle\Big)\\
    &+c_2\Big(\left|5_{x3}\right\rangle+\left|\hA_{x3}\right\rangle\Big)
    +c_3\Big(\left|9_{x3}\right\rangle+\left|6_{x3}\right\rangle\Big)
  \end{split}
\end{equation}
in fact,
\begin{equation}
  \label{eq:52}
  \begin{split}
    \left|0_{z3}\right\rangle +\left|\hF_{z3}\right\rangle=&\frac12\Big[
    \Big(\left|0_{x3}\right\rangle+\left|\hF_{x3}\right\rangle\Big)
    +\Big(\left|3_{x3}\right\rangle+\left|\hC_{x3}\right\rangle\Big)\\
    &+\Big(\left|5_{x3}\right\rangle+\left|\hA_{x3}\right\rangle\Big)
    +\Big(\left|9_{x3}\right\rangle+\left|6_{x3}\right\rangle\Big)\Big].
  \end{split}
\end{equation}
In other words, there is no possibility to combine the value 0 or
$\hF$ in one basis (and no spin sum zero results there) with
\emph{only} spin sum zero in the other. And this property of the
quantum state space together with requirements (i)--(iv) makes it
impossible for Eve to combine an extreme spin sum result at one site
at one setting with a spin sum zero at the other site at the other
setting.

Thus, Alice and Bob \emph{absolutely must} augment their test for
noise in the key with a

\begin{quote}
  Test to see whether the full local results at one setting is
  independent of the spin sum at the remote site at the other setting.
\end{quote}
 
Testing independence of \emph{the spin sums} at different settings is
\emph{not enough}. I would perhaps go so far as conjecturing that the
protocol would be secure given the mentioned two tests (and a test
that the local results occur with equal probability), but I would not
really recommend using them as tests of security of the protocol.
There are two reasons for this: they would provide a relatively weak
statistical test, since the test cases occur quite rarely; but a more
important reason is that the tests are complicated to motivate, and it
would perhaps be difficult to convince a potential user that it
ensures security, even if it may be possible to present a formal
security proof.  The lesson from this is instead that one needs to
analyze \emph{the full local data set} together with \emph{information
  from the remote site}. Since this is necessary, one may as well
choose a much simpler and more easily motivated security test that
does respect this structure:

\begin{quote}
  Test the individual qubits at the two sites. That is, test whether
  the local results at one setting is the same as the remote results
  at the same setting.
\end{quote}

This test would provide the kind of security intended in
\cite{HomeWhit}, and would fail if Eve uses the kind of source
described above.

\section{Conclusions}

\label{sec:conclusions}

The communication protocol proposed in \cite{HomeWhit} does clearly
not use any specifically \emph{quantum} properties of the quantum
teleportation setup. In particular it does not need an entangled state
since a separable state performs equally well. The system is using
classical postselection of appropriate experimental runs to transfer
the data. One could have hoped that the \emph{quantum} properties of
the quantum system of either Section~\ref{sec:introduction} or
\ref{sec:separ-state-impl} would enhance Bob's chance of
distinguishing Alice's setting (a) from (b), but this is not the case,
since the protocol shows the same performance using the purely
classical system of Section~\ref{sec:coin-toss-impl}.

As to the quantum-cryptographic protocol of \cite{HomeWhit}, the usage
of several qubits to transmit a single key bit is problematic. Simply
testing for noise in the sifted key is not sufficient in this
quantum-cryptographic scheme, because Eve can use entanglement to her
benefit, to eavesdrop without being noticed. In
Section~\ref{sec:perfect-illusion} we noted that an additional test
was needed; a test for independence of the remote spin sum with the
full local result. But a simpler test was also suggested; to test
whether the remote qubit results are identical to the local qubit
results if the same setting is used at the two sites.

Of course, Eve will have a difficult time establishing the entangled
mixed state needed for eavesdropping when Alice and Bob use the
originally proposed test. But this is more a technological issue; the
above reasoning is talking about the protocol as being (in)secure
\emph{in principle}, just as \cite{HomeWhit} is talking about the
protocol as being usable \emph{in principle}.

Since it is necessary to test the individual qubits to obtain a secure
system, the protocol does seem wasteful because it only uses
entanglement present in groups of $4N$ qubits for key transmission.
Also, when Alice's raw key bit is 1, no entanglement at all is used,
since the behaviour of the protocol in this situation is derived from
the fact that Bob's results are statistically independent from
Alice's. Many runs will also be discarded because Alice will send
``OK'' quite seldom, since criterion Q (see
Section~\ref{sec:quantum-cryptography}) will be fulfilled with
probability only $\tfrac6{64}$.

In all, little of the available entanglement is put to good use, even
when the qubits are individually tested, since the sifted key is
derived from a joint result of several qubit measurements.
Entanglement is a valuable resource, and should be used with care. In
conlusion, while these protocols are theoretically interesting, they
are probably not very useful in practice.


\begin{thebibliography}{16}
\expandafter\ifx\csname natexlab\endcsname\relax\def\natexlab#1{#1}\fi
\expandafter\ifx\csname bibnamefont\endcsname\relax
  \def\bibnamefont#1{#1}\fi
\expandafter\ifx\csname bibfnamefont\endcsname\relax
  \def\bibfnamefont#1{#1}\fi
\expandafter\ifx\csname citenamefont\endcsname\relax
  \def\citenamefont#1{#1}\fi
\expandafter\ifx\csname url\endcsname\relax
  \def\url#1{\texttt{#1}}\fi
\expandafter\ifx\csname urlprefix\endcsname\relax\def\urlprefix{URL }\fi
\providecommand{\bibinfo}[2]{#2}
\providecommand{\eprint}[2][]{\url{#2}}

\bibitem[{\citenamefont{Bennett et~al.}(1993)\citenamefont{Bennett, Brassard,
  Cr\'epeau, Jozsa, Peres, and Wootters}}]{BBCJPW}
\bibinfo{author}{\bibfnamefont{C.~H.} \bibnamefont{Bennett}},
  \bibinfo{author}{\bibfnamefont{G.}~\bibnamefont{Brassard}},
  \bibinfo{author}{\bibfnamefont{C.}~\bibnamefont{Cr\'epeau}},
  \bibinfo{author}{\bibfnamefont{R.}~\bibnamefont{Jozsa}},
  \bibinfo{author}{\bibfnamefont{A.}~\bibnamefont{Peres}}, \bibnamefont{and}
  \bibinfo{author}{\bibfnamefont{W.~K.} \bibnamefont{Wootters}},
  \bibinfo{journal}{Phys. Rev. Lett.} \textbf{\bibinfo{volume}{70}},
  \bibinfo{pages}{1895} (\bibinfo{year}{1993}).

\bibitem[{\citenamefont{Bouwmeester et~al.}(1997)\citenamefont{Bouwmeester,
  Pan, Mattle, Eibl, Weinfurter, and Zeilinger}}]{BPMEWZ}
\bibinfo{author}{\bibfnamefont{D.}~\bibnamefont{Bouwmeester}},
  \bibinfo{author}{\bibfnamefont{J.-W.} \bibnamefont{Pan}},
  \bibinfo{author}{\bibfnamefont{K.}~\bibnamefont{Mattle}},
  \bibinfo{author}{\bibfnamefont{M.}~\bibnamefont{Eibl}},
  \bibinfo{author}{\bibfnamefont{H.}~\bibnamefont{Weinfurter}},
  \bibnamefont{and}
  \bibinfo{author}{\bibfnamefont{A.}~\bibnamefont{Zeilinger}},
  \bibinfo{journal}{Nature (London)} \textbf{\bibinfo{volume}{390}},
  \bibinfo{pages}{575 } (\bibinfo{year}{1997}).

\bibitem[{\citenamefont{Boschi et~al.}(1998)\citenamefont{Boschi, Branca,
  Martini, Hardy, and Popescu}}]{boschi98:_exper_einst_podol_rosen}
\bibinfo{author}{\bibfnamefont{D.}~\bibnamefont{Boschi}},
  \bibinfo{author}{\bibfnamefont{S.}~\bibnamefont{Branca}},
  \bibinfo{author}{\bibfnamefont{F.~D.} \bibnamefont{Martini}},
  \bibinfo{author}{\bibfnamefont{L.}~\bibnamefont{Hardy}}, \bibnamefont{and}
  \bibinfo{author}{\bibfnamefont{S.}~\bibnamefont{Popescu}},
  \bibinfo{journal}{Phys. Rev. Lett.} \textbf{\bibinfo{volume}{80}},
  \bibinfo{pages}{1121} (\bibinfo{year}{1998}).

\bibitem[{\citenamefont{Nielsen et~al.}(1998)\citenamefont{Nielsen, Knill, and
  Laflamme}}]{NKL}
\bibinfo{author}{\bibfnamefont{M.~A.} \bibnamefont{Nielsen}},
  \bibinfo{author}{\bibfnamefont{E.}~\bibnamefont{Knill}}, \bibnamefont{and}
  \bibinfo{author}{\bibfnamefont{R.}~\bibnamefont{Laflamme}},
  \bibinfo{journal}{Nature (London)} \textbf{\bibinfo{volume}{396}},
  \bibinfo{pages}{52 } (\bibinfo{year}{1998}).

\bibitem[{\citenamefont{Koniorczyk et~al.}(2001)\citenamefont{Koniorczyk, Kiss,
  and Janzky}}]{KKJ}
\bibinfo{author}{\bibfnamefont{M.}~\bibnamefont{Koniorczyk}},
  \bibinfo{author}{\bibfnamefont{T.}~\bibnamefont{Kiss}}, \bibnamefont{and}
  \bibinfo{author}{\bibfnamefont{J.}~\bibnamefont{Janzky}},
  \bibinfo{journal}{J. Phys. A: Math. Gen.} \textbf{\bibinfo{volume}{34}},
  \bibinfo{pages}{6949} (\bibinfo{year}{2001}).

\bibitem[{\citenamefont{Peres}()}]{quant-ph/0304158}
\bibinfo{author}{\bibfnamefont{A.}~\bibnamefont{Peres}},
  \bibinfo{note}{quant-ph/0304158}.

\bibitem[{\citenamefont{Popescu}(1994)}]{popescu94}
\bibinfo{author}{\bibfnamefont{S.}~\bibnamefont{Popescu}},
  \bibinfo{journal}{Phys. Rev. Lett.} \textbf{\bibinfo{volume}{72}},
  \bibinfo{pages}{797} (\bibinfo{year}{1994}).

\bibitem[{\citenamefont{Gisin}(1996)}]{gisin96:_nonloc}
\bibinfo{author}{\bibfnamefont{N.}~\bibnamefont{Gisin}},
  \bibinfo{journal}{Physics Letters A} \textbf{\bibinfo{volume}{210}},
  \bibinfo{pages}{157} (\bibinfo{year}{1996}).

\bibitem[{\citenamefont{Horodecki et~al.}(1999)\citenamefont{Horodecki,
  Horodecki, and Horodecki}}]{horodecki99:_gener}
\bibinfo{author}{\bibfnamefont{M.}~\bibnamefont{Horodecki}},
  \bibinfo{author}{\bibfnamefont{P.}~\bibnamefont{Horodecki}},
  \bibnamefont{and}
  \bibinfo{author}{\bibfnamefont{R.}~\bibnamefont{Horodecki}},
  \bibinfo{journal}{Phys. Rev.~A} \textbf{\bibinfo{volume}{60}},
  \bibinfo{pages}{1888} (\bibinfo{year}{1999}).

\bibitem[{\citenamefont{\.Zukowski}(2000)}]{zukowski00:_bell}
\bibinfo{author}{\bibfnamefont{M.}~\bibnamefont{\.Zukowski}},
  \bibinfo{journal}{Phys. Rev.~A} \textbf{\bibinfo{volume}{62}},
  \bibinfo{pages}{032101} (\bibinfo{year}{2000}).

\bibitem[{\citenamefont{Clifton and Pope}(2001)}]{clifton01}
\bibinfo{author}{\bibfnamefont{R.}~\bibnamefont{Clifton}} \bibnamefont{and}
  \bibinfo{author}{\bibfnamefont{D.}~\bibnamefont{Pope}},
  \bibinfo{journal}{Physics Letters A} \textbf{\bibinfo{volume}{292}},
  \bibinfo{pages}{1} (\bibinfo{year}{2001}).

\bibitem[{\citenamefont{Bennett and Brassard}(1984)}]{BB84}
\bibinfo{author}{\bibfnamefont{C.~H.} \bibnamefont{Bennett}} \bibnamefont{and}
  \bibinfo{author}{\bibfnamefont{G.}~\bibnamefont{Brassard}}, in
  \emph{\bibinfo{booktitle}{Proc. of the IEEE Int. Conf. on Computers, Systems,
  and Signal Processing, Bangalore, India}} (\bibinfo{organization}{IEEE},
  \bibinfo{address}{New York}, \bibinfo{year}{1984}), pp.
  \bibinfo{pages}{175--179}.

\bibitem[{\citenamefont{Ekert}(1991)}]{Ekert91}
\bibinfo{author}{\bibfnamefont{A.~K.} \bibnamefont{Ekert}},
  \bibinfo{journal}{Phys. Rev. Lett.} \textbf{\bibinfo{volume}{67}},
  \bibinfo{pages}{661} (\bibinfo{year}{1991}).

\bibitem[{\citenamefont{Gisin et~al.}(2002)\citenamefont{Gisin, Ribordy,
  Tittel, and Zbinden}}]{GRTZ}
\bibinfo{author}{\bibfnamefont{N.}~\bibnamefont{Gisin}},
  \bibinfo{author}{\bibfnamefont{G.}~\bibnamefont{Ribordy}},
  \bibinfo{author}{\bibfnamefont{W.}~\bibnamefont{Tittel}}, \bibnamefont{and}
  \bibinfo{author}{\bibfnamefont{H.}~\bibnamefont{Zbinden}},
  \bibinfo{journal}{Rev. Mod. Phys.} \textbf{\bibinfo{volume}{74}},
  \bibinfo{pages}{145} (\bibinfo{year}{2002}).

\bibitem[{\citenamefont{Home and Whitaker}(2003)}]{HomeWhit}
\bibinfo{author}{\bibfnamefont{D.}~\bibnamefont{Home}} \bibnamefont{and}
  \bibinfo{author}{\bibfnamefont{M.~A.~B.} \bibnamefont{Whitaker}},
  \bibinfo{journal}{Phys. Rev.~A} \textbf{\bibinfo{volume}{67}},
  \bibinfo{pages}{022306} (\bibinfo{year}{2003}).

\bibitem[{\citenamefont{Aerts et~al.}(1999)\citenamefont{Aerts, Kwiat, Larsson,
  and {\.Z}ukowski}}]{Jalar99c}
\bibinfo{author}{\bibfnamefont{S.}~\bibnamefont{Aerts}},
  \bibinfo{author}{\bibfnamefont{P.}~\bibnamefont{Kwiat}},
  \bibinfo{author}{\bibfnamefont{J.-{\AA}.} \bibnamefont{Larsson}},
  \bibnamefont{and}
  \bibinfo{author}{\bibfnamefont{M.}~\bibnamefont{{\.Z}ukowski}},
  \bibinfo{journal}{Phys. Rev. Lett.} \textbf{\bibinfo{volume}{83}},
  \bibinfo{pages}{2872} (\bibinfo{year}{1999}).

\end{thebibliography}
\newcommand{\sortnoop}[1]{}

\end{document}